\documentclass[aps,prl,twocolumn,groupedaddress,showpacs,amsmath,amssymb]{revtex4}
\usepackage{epsfig}
\usepackage{amssymb}
\usepackage{amsmath}
\usepackage{graphicx}
\usepackage{amsfonts}
\usepackage{epsfig}
\usepackage{revsymb}
\usepackage{bm}
\usepackage{latexsym}
\usepackage{hyperref}

\begin{document}

\title{Observation of a jump-like transition from fast to normal kinetics of facets for free-growing helium crystals.}

\author{V. L. Tsymbalenko}

\email[]{vlt49@yandex.ru}
\affiliation{Kurchatov Institute, 123182 sq.Kurchatov 1 \\
                  Kapitza Institute for Physical Problems RAN, 119334 Kosigin st.2 \\ Moscow, Russia}


\begin{abstract}
The growth kinetics is studied of free-growing helium crystals within the temperature range 0.5-0.83K and supersaturations to $\sim$10mbar. For the pressure supersaturations above $\sim$0.5mbar, the numerous drastic changes are observed in the growth kinetics for all crystalline facets from the normal and slow kinetics to anomalously high or burst-like one.  It is found for the first time that the anomalous-to-normal transition in the facet growth kinetics occurs in a jump-like manner within the time 80$\mu$s interval of the measurement record.

\end{abstract}


\maketitle

\section{Introduction}\label{sec1}

The theoretical paper by Andreev and Parshin in 1978 \cite{AP} has aroused  experimental interest in the growth kinetics of $^4$He crystals. They have suggested that the quantum nature of $^4$He crystals should result in delocalizing the elementary defects as crystal facet steps in the surface structure. Above the roughening transition temperature the free energy of crystal steps vanishes and thus their equilibrium concentration is high. The deviation from the phase equilibrium measured in units of pressure $\delta p$ sets the crystal steps in motion limited by the small viscosity of phonon gas. Accordingly, the large mobility of the crystal steps results in a high mobility of the crystalline surface. The mobility of crystalline surface increases as the temperature  lowers. This theoretical prediction has received the experimental confirmation resulting in a wide theoretical and experimental study of the effect, see review~\cite{BPH}.

Below the roughening transition temperature the free energy of crystal steps is positive, leading to appearance of flat segments on the equilibrium crystal, namely, facets. The steps on the crystal facets are delocalized and have high mobility. The thermodynamically equilibrium step concentration  decreases  exponentially with decreasing the temperature.  This should lead to the complete loss of the facet mobility. The facet growth in this temperature range is provided by the topological sources of the steps. Termination of a single screw dislocation to the surface generates a spiral facet growth. A pair of screw dislocations with opposite sign represents the Frank-Reed source. The concentration of such sources depends on the crystal defects different at the equivalent facets. As a result, the free-growing crystals have an irregular hexagonal shape \cite{VLT1995}. The facet growth rates measured experimentally have demonstrated the specific features inherent in the indicated mechanisms typical also for classical crystals. The growth of crystal facets starts above the threshold supersaturation $p_ {FR}$ and the growth rate depends nonlinearly on the magnitude of the deviation from phase equilibrium $\delta p$. For small deviations from phase equilibrium, the facet growth kinetics of quantum crystal does not differ in kind from that of classical crystal facets. For this reason, the interest in such studies was low.

The situation changes drastically as a result of studying the crystal facet growth by the  Finnish-Russian group in the 2-250 mK range~\cite{RF-1}. In a number of experiments it was possible to grow the defect-less crystals. For the above-mentioned temperatures, the facet is free of thermodynamically equilibrium steps and, as is discussed previously, the facet should not grow. However, in the experiments under constant helium flow into the container it is found that the pressure increases monotonically until the critical supersaturation $\delta p_c$ is reached after which a drastic drop has been observed. This means that the ideal facet does not grow until qualitatively new growth mechanism of unknown nature starts in operation. This provides the facet growth for some time sufficient for crystallizing an excess mass of the supersaturated liquid and decreasing the pressure close to the phase equilibrium one. Then this mechanism disappears, the facet looses its mobility and the process repeats.

The similar phenomenon is discovered in the same year in our experiments on the crystal growth  from the metastable liquid at temperature 0.44 K~\cite{VLT1996}. The crystal growth time at initial supersaturation $\delta p_0 \sim 14$mbar is shorter than 0.7ms, corresponding to the growth rate higher 1~m/s. This magnitude is much higher than the growth rate extrapolated to 14mbar from the measurements carried out at $\delta p \sim1$~mbar \cite{VLT1995}. Formally introducing the kinetic growth coefficient $K$ as the coefficient of proportionality between the average growth rate of the facet and the difference in the chemical potentials of the crystal and the liquid, it is possible to quantify the change in kinetics in this process. The rapid growth results from the growth $K$ coefficient enhancement by two or three orders of magnitude as compared with those measured for the crystals free from such  transition~\cite{VLT1995}. The further studies show that this effect has many common features with the phenomenon observed in Helsinki. The temperature dependences $\delta p_{c}(T)$ are consistent.  The nature of small $^3$He impurity  effect on the critical supersaturation is similar in kind. The review of studies on abnormal growth (burst-like growth) is given in paper~\cite{UFN}.

The conditions are studied in detail for the transition of crystal facets to the anomalously fast growth. The reverse process of restoring their normal kinetics is studied, in fact, only in a single  work on the free-growing helium crystals~\cite{VLT2004}. After $0.02-0.1$ s the growth rate of the facets returns to the magnitudes typical for the normal crystals. According to~\cite{RF-1}, one can only estimate the time of returning to the growth-free state after the pressure drop. This time is determined by the pressure measurement time constant $\sim 0.5$ s. After this time, the facet becomes motionless again.

The change in the growth kinetics in the 0 -- 0.02 s range, where the main relaxation of the facet kinetics occurs, remains unknown. To solve this problem, the crystal growth technique is modified~\cite{VLT2021}. After nucleation in a metastable liquid, the crystal continues to grow under constant flow of helium into the container. In these experiments the liquid helium flow exceeds significantly that in our previous experiments. It turns out that the pressure records show the numerous jumps similar to those observed in~\cite{RF-1}. In other words, new technique makes it possible to observe numerous transitions of crystal facets from the normal slow growth to anomalously fast growth with the high time resolution. In this work we continue to study the return of the crystal facet kinetics to the normal one after the anomalously fast growth stage.

\section{Experimental Method}\label{sec2}

The crystals are grown in the opaque container as described earlier, see \cite{UFN}. The inner volume of the container is 330 mm$^3$. The maximum initial pressure supersaturation is observed during the crystal nucleation on the container wall and achieves 23 mbar. The reduction of the starting pressure to 0.5--15 mbar is obtained by applying the high voltage to the tungsten needle located at the center of the container. The distance from the needle tip to the inner container walls exceeds 1.1 mm. Accordingly, the crystals grow freely without touching the walls. One of the container walls operates as a membrane of capacitive sensor with the resonance frequency 13.6 kHz. The pressure in the container $p_1(t)$ is recorded with step 80 $\mu$s.

At the end of crystal growth when the facet kinetic growth coefficient becomes sufficiently large, the pressure oscillations are observed on the wall of the capacitive sensor. Their period, estimated from the container volume for the starting supersaturations 2--23 mbar, lies in the range 120--190 $\mu$s \cite{VLT2000}. The registration of oscillations performed in increments 80 $\mu$s does not allow us to determine accurately the shape of dependence $p_1(t)$ and to calculate the magnitude of the average growth coefficient $K$ at the fast growth stage according to the method in~\cite{VLT2001}. Note that the observation of pressure oscillations evidences for the high magnitude of growth coefficient $K$.

The crystal nucleation and its subsequent growth occur in the course of the continuous flow of a liquid into the container. This is a consequence of the decrease in external volume $V_2$ due to the compression of the bellows by the engine. After nucleating the crystal and its growth, the direction of the motor rotation is reversed, the volume of the external system increases, the flow of helium comes from the container, and the crystal size reduces until it completely melts.

The crystal volume is calculated from the pressure-time dependences $p_1(t)$ in the container. The crystals grow and melt under conserving the total mass of helium in the container $M_1$ and the mass of gas $M_2$ in the external pressure system. So, we have
\begin{align}
&M_1+M_2 = const, \nonumber\\
&M_1 = \rho' V_c +\rho (p_1) (V_1 - V_c),\nonumber\\
&M_2 = p_2 V_2(t) \frac{\mu_4}{R_{gas} T},\nonumber\\
&\dot{M_1}=-\dot{M_2}=\frac{p_2 - p_1}{Z}.
\label{Eq1}
\end{align}
Index 1 refers to the parameters in the container and index 2 does to the outer part.
Here $p$ is the pressure, $V_c$ is the volume of the crystal, $V_1$ is the internal volume of the container, $V_2$ is the internal volume of the external pressure system at room temperature, $Z$ is the impedance of the inlet capillary, $\rho$ and $\rho'$ are the densities of liquid and solid helium, respectively, $T$ is the gas temperature in outer pressure system, $\mu_4$~=~4.0026~gm/mole and $R_{gas}$ = 8.314~J/mole*K.
The compressibility of helium gas at 25 bar differs from that of an ideal gas by 1.1\%. The third equation for the mass of gas $M_2$ is written with such accuracy. In these experiments, the ratio of the volume of the crystal grown at an initial supersaturation $\delta p_0 \sim$20mbar to the internal volume of the container is $V_c/V_1 = \rho /\Delta\rho k_l \delta p_0 \sim 10^{-3} \ll 1$. Taking this fact into account and transforming the equations, we get
\begin{align}
& \frac{dp_1}{dt}+ \frac{\Delta\rho}{\rho}\frac{1}{k_L V_1}\frac{dV_c}{dt} = \frac{p_2 - p_1}{\tau_1}, \nonumber\\
& \frac{dp_2}{dt}=-\frac{p_2 - p_1}{\tau_2} +\left(1+\frac{\tau_1}{\tau_2}\right) \dot p_{10},\nonumber\\
&\tau_1=Z \rho V_1 k_L, \tau_2 = Z \frac{\mu_4 V_2}{R_{gas} T}
\label{Eq2}
\end{align}
where $k_L$ is the liquid helium compressibility and $\Delta \rho =\rho'-\rho$. The rate of increasing the fluid pressure in the container $\dot {p_1} =$ const $= \dot p_{10} >0$ is determined by the rate of changing the volume of the external system $\dot {V_2}<0$, remaining constant in the course of the crystal nucleation and growth and reversing the sign as the crystal starts melting.  This gives us the initial conditions for the pressures at the moment of crystal nucleation
\begin{align}
p_1(0) = \delta p_0, p_2(0) = \delta p_0 + \tau_1 \dot p_{10}.
\label{Eq3}
\end{align}
Integrating the second equation in Eq. (\ref {Eq2}) yields us the pressure variation $p_2(t)$ in the external system
\begin{multline}
p_2(t)=\frac{1}{\tau_2}\int\limits_0^t p_1(x)\exp\left(\frac{x-t}{\tau_2}\right)dx+ \\
+(\tau_1+\tau_2)\left[1-\exp\left(\frac{-t}{\tau_2}\right)\right] \dot p_{10}
+p_2(0)\exp\left(\frac{-t}{\tau_2}\right) .
\label{Eq4}
\end{multline}
Integrating the first equation (\ref {Eq2}), we obtain the temporal dependence of the crystal volume
\begin{align}
V_c(t) = \frac{\rho}{\Delta\rho} k_L V_1 \left[p_1(0) -p_1(t) + \frac{1}{\tau_1} \int\limits_0^t (p_2 - p_1) dt' \right ].
\label{Eq5}
\end{align}
Note that formulas (4) and (5) are derived under the assumption that the impedance $Z$ is constant during crystal growth. However, this assumption has not been experimentally confirmed. For this reason, we do not calculate the dependence of the crystal volume on time in this work.

A ratio of the time constants is independent of the impedance of the capillary and  determined by a ratio of the helium masses in the container and in its outer part
\begin{align}
\frac{\tau_1}{\tau_2}=\frac{M_1}{M_2} k_L p_{phase} = 0.069.
\label{Eq6}
\end{align}
Here $p_{phase} \cong 25$ atm is the phase equilibrium pressure. The time constant
$\tau_{1} \cong$ is 8.2 ms and $\tau_{2} \cong$ is 120ms. This is much less as compared with  the time constants in \cite{VLT2004} where these constants are 37s and 30s, respectively.

In the above equations the pressure $p_1$ is measured from the phase equilibrium pressure for the flat interface. In the experiment the similar situation is not realized due to the curvature of atomic rough regions of the crystal surface and thus the pressure is always higher. Its smallest distinction from the phase equilibrium pressure takes place at the moment of starting the crystal melting. The crystal volume is maximal and the crystal facets disappear as can be seen in Fig.2.  The correction to the pressure can approximately be estimated with calculating the curvature radius of the crystal volume. In these experiments the correction to the pressure is about 0.05--0.1 mbar.

As is noted in the Introduction, the dependence of the crystal facet growth rate on the supersaturation demonstrates a threshold different for the \textit{c}- and \textit{a}-facets. Even the equivalent crystal facets of a free-growing crystal have the various growth rates. For these reasons, it is impossible to determine such dependences for each crystal facet by measuring $p_1(t)$ alone. We will use the kinetic growth coefficient $K$ averaged over the whole surface. The dependence of surface growth rate $V_{surf}$ on the supersaturation, i.e., difference in the chemical potentials of the phases $\Delta \mu$, is approximated by the linear function. The pressure supersaturation of the crystal-liquid phase diagram in this region is determined by the pressure difference in the container from the phase equilibrium pressure with the correction for the curvature $R$ of the atomic rough surface segments (edges).
\begin{equation}
V_{surf} = K \left(\frac{\Delta\rho}{\rho\rho'} p_1 - \frac{\alpha}{\rho' R}\right).
\label{Eq7}
\end{equation}
Here $\rho$ and $\rho'$ are the densities of liquid and solid helium, $\Delta \rho = \rho' - \rho> 0 $, and $\alpha $ is the surface energy assumed to be isotropic.

Under the constant flow of liquid into the container a drastic change in the crystal facet kinetics is seen from the jump-like singularity of the derivative in the $p_1(t)$ records. From the experimental data it follows that the transition from the slow to the fast kinetics occurs faster than 40 $\mu$s \cite{UFN}. The reverse transition, as it follows from the experimental results presented below, takes place faster than the measurement step 80 $\mu$s. For this time, the rate of increasing the helium mass in the container and the crystal surface area $S_{cryst}$ vary  negligibly. The incoming mass increases that of the crystal. We have
\begin{equation}
\dot{M_1} \cong const \cong \Delta\rho S_{cryst} V_{surf} \cong \frac{\Delta\rho^2}{\rho\rho'} S_{cryst} K p_1 \sim  K p_1.
\label{Eq8}
\end{equation}
We get that in the stationary case, $p_1 \sim 1/K$. After the fast-to slow transition of kinetics, the supersaturation changes from small magnitude to large one with the time constant $\tau_1$, see the first equation of system (\ref{Eq2}).

The kink in the derivative from almost zero magnitude to the finite one means that all the possibilities of rapid growth, i.e., with the \textit{c}- and \textit{a}-facets, are restricted. Note that, in order to stop the crystal growth, it is sufficient to lose the mobility of three \textit{a}-facets at angle $120^\circ$. If the intermediate crystal facets continue to grow rapidly, their sizes will reduce rapidly and the crystal shape will become closer to the triangular prism. To summarize, the jump in the derivative of the $p_1(t)$ dependence from the small to large magnitudes evidences unambiguously for blocking the crystal growth with the \textit{c}-facets and at least with three  \textit{a}-facets located at angle $120^\circ$.

\section{Experimental results}
\subsection{Types of behavior $p_1(t)$ observed}
In experiments on crystal growth in a metastable liquid\cite{UFN}, two types of the pressure curve  behavior are observed after the crystal nucleation. In the both cases the pressure drops to the phase equilibrium one but this occurs at very different times. The crystals demonstrating the normal growth grow slowly and for a long time. The  facet kinetics remains  unchanged in time and does not differ from the kinetics of the  crystal facets grown at low supersaturations. The drastic drop in the pressure for the short time means the transition of the crystal facets to the rapidly growing state like burst-like growth. The start of the burst-like growth regime occurs with some delay with respect to the moment of crystal nucleation. The higher the initial supersaturation $\delta p_0 = p_1(0)$ of the metastable liquid, the shorter the delay time.  Provided the  facet kinetics is high, the excitation of the low-frequency damped pressure oscillations is observed in the container. As the pressure decreases to the phase equilibrium one for 20--100 ms, the facet kinetics returns to the magnitudes typical for the normal slow growth of crystals.
\begin{figure}
\begin{center}
\includegraphics[%
  width=0.65\linewidth,
  keepaspectratio]{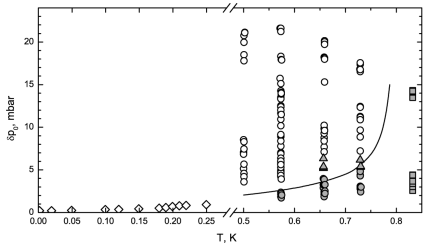}
\end{center}
\caption{The crystals selected by the growth type dependent on supersaturation $\delta p_0$ at  which the crystal nucleates. The solid curve separates the regions of normal and abnormal growth described earlier in review \cite{UFN}. Experimental results are shown by symbols. The grey squares refer to the normal growth crystals, see Section~3.1. The grey circles are the crystals grown normally with the pressure splashes in the growth process. The triangles are the crystals for which the pressure $p_1$ jumps are observed after the normal growth stage. The empty circles indicate the crystals showing the rapid growth and oscillation after drastic pressure drop. The diamonds show the result of the article\cite{RF-1}.}
\label{fig1}
\end{figure}{}

Figure~1 shows the boundary between the crystals that have the features in the $p_1(t)$ record characteristic for the burst-like growth regime and the crystals that have no such features in a set of previous experimental data \cite{UFN}. The same figure shows the symbols corresponding to the temperature and supersaturation $\delta p_0$ at the moment of crystal nucleation according to the results of this work. It can be seen that the separation of the $T - \delta p_0$ diagram into the normal and abnormal growth regions is conserved. This Section shows the measurement results for the crystals grown on a tip and having no contact with the container wall. For this reason, we restrict ourselves to recording $p_1(t)$ for 0.3--0.5 s after the crystal nucleation.

Note that our classification of crystal growth process by time into two stages, as is discussed in Sec. 3.1, is most suitable for the burst-like growth mode. After the crystal facets acquire the fast growth kinetics, the crystal growth occurs at almost constant helium mass in the container. The further crystal growth and the supersaturation magnitude are now governed by the crystal facet kinetics and the liquid inflow from the external system under increasing the helium mass in the container.

\subsection{Growth of faceted crystals that do not exhibit the burst-like growth mode.  T=0.83 K}

An example of the pressure change in the container after crystal nucleation is shown in Fig. 2. Several characteristic segments are visible on the record. After nucleating the crystal, the pressure drops rapidly over $\sim$3 ms, section A-B. The crystal growth occurs at almost constant mass of helium in the container. This simplifies the system of equations (\ref{Eq2}). The average kinetic growth coefficient is ${K\approx2.3\cdot10^{-3}}$ s/cm. This value is in agreement with the data obtained previously for the free-growing crystals, see Fig. 4 \cite{UFN}.

It is seen that the pressure does not drop to the phase equilibrium pressure. The reason lies in its small magnitude. It can be seen from relation (\ref{Eq7}) that the crystal growth continues as long as the pressure in the liquid exceeds the contribution from the curvature of the edges. In the previous experiments in the containers with internal volume 10--30 times larger this effect is negligible. The growth of a spherical crystal at various initial supersaturation taking into account the curvature of the surface is considered in Appendix I. Figure 10 shows the experimental data on the pressure $p_{stop}$, at which the growth of the crystal stops at point B. The experimental points lie above the model calculation. This means that the shape of the crystals differs from the spherical one.

The crystal growth regime changes at the B-C-C' site. The quasi-stationary growth continues due to the liquid flow into the container. The supersaturation is determined by the balance between the fluid flow and the crystallization rate. As the crystal grows, the surface area of the crystal increases, the pressure in the external system decreases with the time constant $\tau_2$, and, as a result, the supersaturation in the container decreases.
\begin{figure}
\begin{center}
\includegraphics[%
  width=0.65\linewidth,
  keepaspectratio]{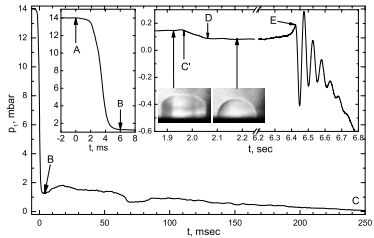}
\end{center}
\caption{Recording of pressure variations during the crystal growth at 0.83 K. The images, obtained in an optical container, illustrate the change in the crystal shape from growth to melting. The right-hand graph in the inset shows the pressure oscillations that occur after the complete melting of the crystal.}
\label{fig2}
\end{figure}

In the interval C'-D, the motor rotation is reversed (Fig.2, middle insert). The outer volume increases, the helium flow is directed out from the container, and the crystal melts. This inset shows the change in the crystal shape from growth to melting. The illustration is obtained in our previous experiments in the  optical container, see \cite{UFN}. During melting, the crystal shape is determined by the most mobile atomically-rough segments of the surface. The edges are significantly reduced and the crystal takes a drop-like shape. Since the curvature of the crystal surface is now less than the curvature of the rounded edges, the pressure drops by the amount corresponding to their difference.

During the melting process, the supersaturation in the container is almost constant, section D-E. Near complete melting, the pressure starts to rise. This is due to increasing the curvature of the crystal surface with decreasing its size, as follows from Eq. (\ref{Eq7}). After disappearing the crystal, low-frequency pressure oscillations at frequency18.9 Hz and decrement 0.66 are observed in the container. The solutions of system (\ref{Eq2}) after crystal melting have a relaxation character. Consequently, in order to obtain the pressure oscillations, the inertial factor must be introduced into the equations. The natural oscillation frequencies of the liquid in the container and the gas in the external system have frequencies much higher than the frequency observed. The only factor unaccounted for in Eqs. (\ref{Eq2}) is the helium mass in the filling capillary connecting the container and the external pressure system. The container-capillary-external volume system is similar to the Helmholtz resonator. The difference is that when the helium flows in the capillary, the restoring force arises from the both sides of the capillary. This model is discussed in Appendix II. The estimate is in the satisfactory agreement with the experimental parameters.

\subsection{Growth of crystals demonstrating the burst-like growth mode}
\subsubsection{T=0.73K}
Figure 3 shows the pressure variation in the container after the crystal nucleation. In the lower record (a) when the initial supersaturation is approximately a half of the boundary magnitude  $\delta p_{c}$, the $p_1(t)$ dependence is similar to the pressure drop during the normal crystal growth, see Fig.2. The kinetic growth rate, estimated from the initial pressure drop, is $K = 2.5\cdot 10^{- 3}$ s/cm. No specific features are observed.
\begin{figure}
\begin{center}
\includegraphics[%
  width=0.65\linewidth,
  keepaspectratio]{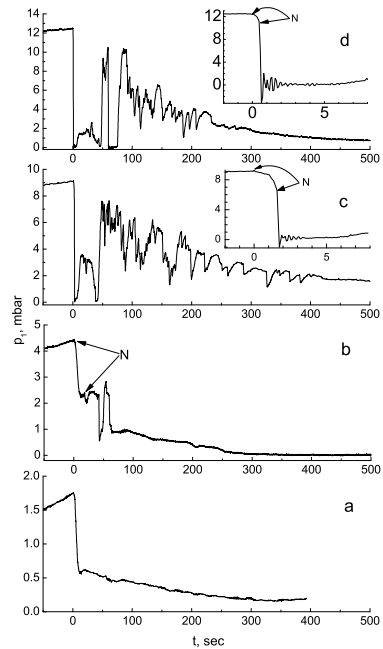}
\end{center}
\caption{The pressure records for the crystal growth at $T=0.73$ K. The symbol N and the arrows indicate the stage of normal crystal growth. }
\label{fig3}
\end{figure}

Plot (b) shows the crystal growth near the boundary separating the normal and fast growth regions. At first, the crystal grows at the kinetic growth rate $K = 1.8\cdot 10^{-3}$ s/cm typical for the normal crystal growth at this temperature. This region is highlighted by the arrows and denoted with symbol N. Later, the pressure starts decreasing, as in the previous plot, and over time $\sim$50 ms  we observe a drastic pressure drop, revival to the pressure before its jump, and the drop once more after which the crystal continues growing without any features like in the previous plot. The drastic pressure drop indicates a rapid acceleration in the crystal surface growth. Then, there is a return to the slow growth kinetics and, after a while, another drastic pressure drop. Eventually, the transition from the normal to burst-like growth is observed twice in the growing crystal at supersaturation $\sim$2~mbar being a half of the boundary supersaturation $\delta p_{c}$ at this temperature.

We give above the records of the pressure for the crystals nucleated in the anomalous region. After nucleation the crystal grows at the normal rate accompanied by the gradual pressure drop. In the inserts these regions are indicated by the arrows. Then the crystal goes over to the burst-like growth state and grows quicker than for 0.6 ms with the pressure drop to the phase equilibrium pressure. The inserts show the rapidly decaying pressure oscillations evidencing for the high magnitude of kinetic growth coefficient.

The plot (c) shows that the pressure increases for the time interval $\sim $35 ms. Later, we have  a pressure drop to the phase equilibrium pressure, succeeded  by  the  next increase in the pressure  to about 80\% of initial supersaturation $\delta p_0$. Then,  there is a monotonous pressure drop with the  concurrent jumps of characteristic shape. The drastic drop in pressure is followed by its rise. When the supersaturation drops to $\sim$2 mbar, the concurrent pressure jumps disappear.

Plot (d) shows the pressure variation during the growth of a crystal nucleated at the starting pressure $\delta p_0\sim12$ mbar. This is similar in kind to the previous plot. Note that the pressure jump appears at $\sim$ 60ms and  similar to that accompanying the crystal nucleation and  exciting the pressure oscillations as well. After  this, for $\sim$12 ms, the pressure supersaturation is small, i.e., the kinetic growth coefficient is large and, correspondingly, the crystal is in an anomalous state during this time interval.

\subsubsection{T=0.66K}
After the crystal nucleation below the critical supersaturation curve ${\delta p_0 < \delta p_{c}(T)}$, a slow growth stage is observed with kinetic coefficient ${K=6.9\cdot 10^{-4}}$ s/cm. This is followed by a monotonous pressure drop, in which we can see one pressure jump due to the transition of the facets to an anomalous rapid growth state. The facet kinetics then returns to the normal slow kinetics. Plot (b) shows the crystal growth near the $\delta p_c (T)$ curve, which proceeds with kinetic growth rate $K = 8.6\cdot 10^{- 4}$ s/cm typical for normal crystal growth at this temperature. The further decrease in pressure occurs with several jumps, indicating the transition of the normal growth regime to the burst-like growth regime and the subsequent restoration of slow kinetics. In the region of anomalous growth (c-d), after the initial jump from the value $\delta p_0$ to the phase equilibrium pressure, numerous pressure jumps are visible. These jumps appear at the supersaturations below $\sim$3 mbar and stop when the pressure drops below $\sim $1mbar.
\begin{figure}
\begin{center}
\includegraphics[%
  width=0.65\linewidth,
  keepaspectratio]{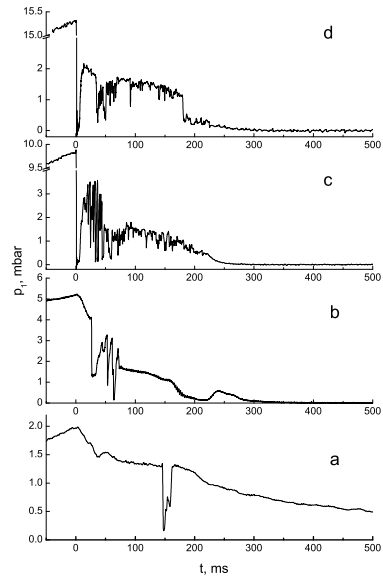}
\end{center}
\caption{The pressure records during the crystal growth at T = 0.66 K.}
\label{fig4}
\end{figure}

\subsubsection{T=0.57 K}
Plot (a) shows the pressure variation accompanying the growth of a crystal nucleated near the boundary  between the normal and anomalous regions. The crystal growth occurs with kinetic coefficient $K = 2.4\cdot10^{- 3}$ s/cm. In $\sim$280 ms after the crystal nucleation at supersaturation of the order of 1mbar the pressure jumps are observed continuing for $\sim$200 ms. For the crystals nucleated in a metastable liquid within the range $\delta p_0 = 1.7-2.4$ mbar,  this is a typical picture. The difference is that the pressure jumps begin to appear on the $p_1(t)$ record with an arbitrary delay relative to the moment of crystal nucleation.
\begin{figure}
\begin{center}
\includegraphics[%
  width=0.65\linewidth,
  keepaspectratio]{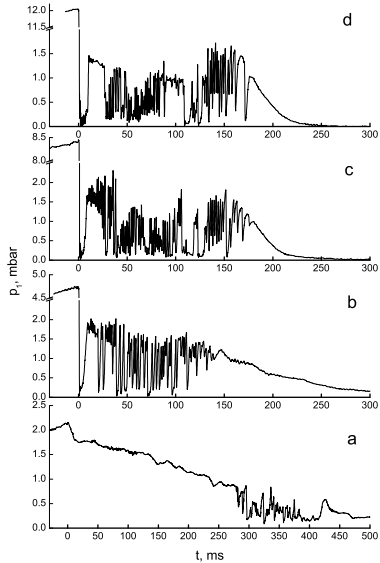}
\end{center}
\caption{The pressure records during the crystal growth at T = 0.57K.}
\label{fig5}
\end{figure}

Graphs (b-d) show the pressure behavior in the container after nucleation and growth of crystals in the burst-like growth mode. The fast growth after the crystal nucleation  results in the damped pressure oscillations. After decay of the oscillations for 5--20 ms, the pressure is close to the phase equilibrium one. Then the supersaturation increases. This leads to numerous pressure surges, reflecting abrupt transitions of the crystal facet kinetics from the burst-like growth regime to the normal slow one. Note that during this process the supersaturation is small as compared with the initial supersaturation $\delta p_0$ and does not exceed $\sim$2 mbar.

\subsubsection{T=0.50K}
The records in Fig.6 clearly show the different modes for the transition of the crystal growth kinetics from normal to anomalous and vice versa. The fast crystal growth after nucleation leads to a sharp drop in pressure. The observed pressure oscillations indicate a high magnitude of kinetic growth coefficient $K$. Taking into account the remarks of Sec. 2, we can determine only the lower boundary for the coefficient according to the methodology of paper \cite {VLT2001}, $K>0.1$ s/cm.
\begin{figure}
\begin{center}
\includegraphics[%
  width=0.65\linewidth,
  keepaspectratio]{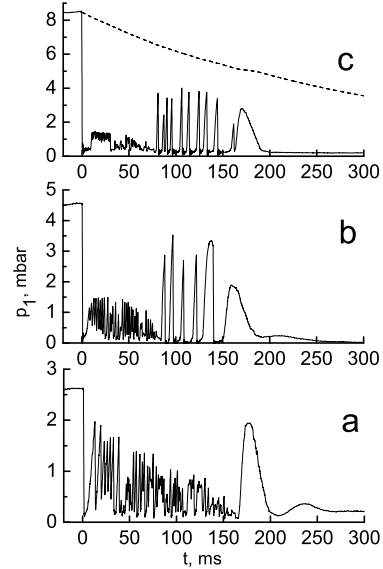}
\end{center}
\caption{The pressure records during the crystal growth at $T = 0.50K$. The dashed line in graph (c) represents the calculated pressure change $p_2(t)$ in the external system.}
\label{fig6}
\end{figure}

At the end of the oscillations, the pressure increases. In plot (a), during the growth of a crystal nucleated at supersaturation 2.6mbar, quasiperiodic pressure jumps with amplitudes $\sim$2 mbar are observed. They decrease in magnitude over time until they stop at $\sim$170 ms after the crystal nucleation. The crystal growth pressure record, shown in graph (b), demonstrates the similar pressure behavior.  The jumps with amplitude of $1.13\pm0.44$ mbar and the interval between them $2.35\pm0.56$ ms are observed for $\sim$80 ms. The supersaturation after the sharp pressure drop is close to zero, $p_{min} = 0.25\pm0.23$ mbar. Then the nature of the process changes. Instead of frequent and numerous jumps with the amplitude not exceeding $\sim$1.5 mbar, 5 jumps are observed with amplitude up to 3.5 mbar. The same picture is observed in the pressure record during the crystal growth nucleated at $\delta p_0 = 8.2$ mbar in the graph (c). The frequent jumps of small amplitude stop after 80 ms after the crystal nucleation. Then ten jumps of large amplitude are visible. During the crystal growth the pressure in the external system drops in spite of the continuing decrease in its volume by the compression of the bellows. The pressure change $p_2(t)$, calculated according to equations (2-5), is shown by the dashed curve.

Record scan (c) is shown in Fig. 7. The time counting on all graphs starts from the moment of crystal nucleation. The bottom graph shows the record of pressure oscillations at the end of the burst-like growth. Let us estimate the averaged kinetic growth coefficient $K$ by the maximum oscillation amplitude. From the calculation of work \cite{VLT2001} it follows that this amplitude changes noticeably from the maximum at $K>1$ s/cm to approximately zero at $K\sim0.03$ s/cm. At values of $K<0.03$ s/cm, oscillations are not excited and the pressure drops monotonically. The initial oscillation amplitude is $\sim$3 mbar, which, according to the method \cite {VLT2001}, gives an estimate as  $K\approx 0.12$ s/cm. This value is consistent with the measurements on for the large crystals, see review \cite{UFN} Fig.4.
\begin{figure}
\begin{center}
\includegraphics[%
  width=0.65\linewidth,
  keepaspectratio]{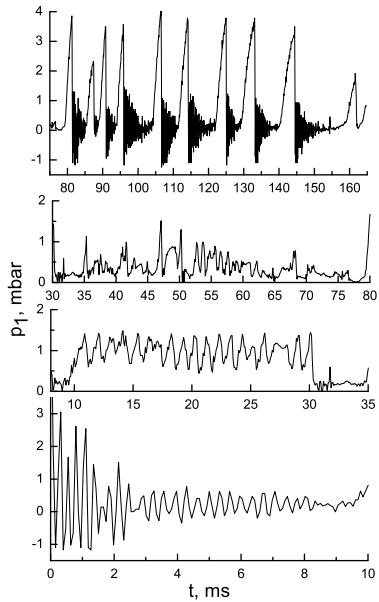}
\end{center}
\caption{The scan recording Fig. 6c from bottom to top: pressure oscillations after crystal nucleation in a metastable liquid and abnormal growth; quasiperiodic pressure surges; chaotic jumps; quasi-periodic jumps of large amplitude.}
\label{fig7}
\end{figure}

The next plot shows the quasi-periodic pressure jumps in the range  10-30 ms. The average interval between them is $1.08\pm 0.24$ ms and the amplitude of the jumps is $0.67\pm 0.16$ mbar. In this mode the overpressure after the jump exceeds noticeably the phase equilibrium  pressure $p_{1 min}=0.66\pm0.15$ mbar. The following plot shows that the nature of the pressure jumps changes later. Within 30-80 ms the jumps occur irregularly, amplitude of several ones reaches 1.5 mbar, and the pressure after the jumps is close to the phase equilibrium pressure.

The next mode is observed within interval 80--170 ms. The crystal is in the state with  the normal growth kinetics for a long time. During this time the pressure in the container increases to magnitudes 2--4 mbar. Then, the jump in the crystal facet growth kinetics from normal to anomalous is observed and accompanied by the rapid crystal growth and exciting the pressure oscillations in the container. The plot shows that the pressure oscillations take place near the phase equilibrium pressure $p_{phase}$.  This means that for this time, the kinetic growth coefficient of the crystal surface remains large and typical for the burst-like growth mode. The average magnitudes of each oscillation period differ from $p_{phase}$ within  $0.04\pm 0.06$ mbar. This means that according to relation (\ref{Eq8}), the kinetic growth coefficient of the crystal surface remains large and indicative in favor of  the burst-like growth mode.

The burst-like growth mode continues for a few milliseconds. Then, at the end of the oscillations, the pressure in the container starts to rise rapidly. As is noted in Sec. 2, a kink in  the time derivative $p_1(t)$ indicates that the \textit{c} facets and at least three \textit{a} facets at  angle $120^\circ$ become inactive. From the width of the derivative jump it is possible to estimate the minimum transition time from the anomalous to normal kinetics for these facets as  about 100$ \mu$s.

The change in the time duration $t_{state}$ of the normal and abnormal state is based on plots 6b, 6c and 7 and is shown in Fig. 8.
\begin{figure}
\begin{center}
\includegraphics[%
  width=0.65\linewidth,
  keepaspectratio]{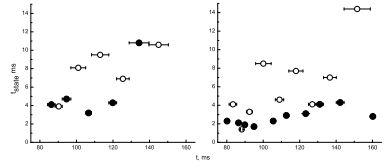}
\end{center}
\caption{The duration $t_{state}$ for  the normal and abnormal state of  the crystal facets during their growth at T = 0.50 K. The left plot shows the data from the record in Fig. 6b for the crystal nucleated at supersaturation 4.5 mbar, Fig. 6b. The right-hand  plot shows the results for the crystal nucleated at the initial supersaturation 8.2 mbar, Figs. 6c and 7. The filled symbols represent the duration for the normal kinetics of the crystal facets. The empty symbols represent the lifetime of the abnormally fast kinetics of the  crystal facets.}
\label{fig8}
\end{figure}
It can be seen that the duration of the normal state kinetics changes negligibly in time and lies within the range 2--4 ms. The lifetime of the abnormal state is longer in average as compared with that of the normal state and increases in time.

\subsection{Anomalies in the kinetics of normal crystal facets.}
The results of subsections 3.3.1--3.3.4 are obtained during a single growth of crystals nucleated in a metastable liquid. To answer the question whether the formation of an anomalous phase in normal crystals is possible, the following experiments are performed. The crystals are nucleated on a tip at the voltage providing the crystal nucleation without noticeable potential barrier. Then the crystal growth stops. After that the crystal melts to the small size. The measurement starts after this procedure. The helium flow into the container is the same as in the previous experiments.
\begin{figure}
\begin{center}
\includegraphics[%
  width=0.65\linewidth,
  keepaspectratio]{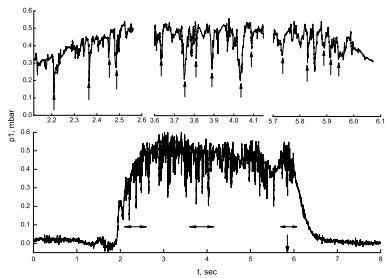}
\end{center}
\caption{The pressure recording during the normal crystal growth at T = 0.66 K. The motor that produces the pressure in the external system starts at the moment $t = 0$. The downward arrow in the lower plot labels the time of the motor reversal and the beginning of the crystal melting. The horizontal arrows denote the time intervals expanded above on the scheme. The upward arrows show the pressure jumps due to the transition of crystal facet growth to the burst-like mode.}
\label{fig9}
\end{figure}

An example of the $p_1(t)$ dependence is shown in Fig.9. It can be seen that no singularities are observed before the pressure supersaturations $\sim0.2$ mbar. In the range 0.2--0.6 mbar the  pressure jumps appear randomly over the record. This shows that when the necessary supersaturation is produced, the facets of  normally grown crystal go over to the burst-like growth state \cite{RF-1,RF-2}.

\section{Discussion}
\subsection{The boundary between the regions of anomalous and normal crystal growth. The temperature effect.}
In Fig. 1 we show the boundary separating the regions of normal and abnormal growth in the $\delta p_{0}-T$ coordinates. This boundary is conditional. The point is that the drastic pressure  drop,  meaning the transition of the  crystal facets to the burst-like growth, is observed some time after the crystal nucleation. Prior to this transition, the crystal grows in the normal manner with a gradual pressure drop in the container. This fact is well described in the review \cite{UFN}, Fig.7--10. Also, it can be seen from the pressure records in this series of experiments (see Fig. 3 and the insets in the  c and d plots). Because of this time delay, the magnitude of the pressure  jump is smaller  than the supersaturation at which the crystal nucleates. If we mark exactly the jump amplitudes in the $\delta p_0(T)$ plot, the boundary moves in the downward direction.

As it follows from the pressure records performed in the range 0.5--0.73 K, the pressure jumps, indicating  the onset of the burst-like growth regime, have the different amplitudes varying
from a few fractions of millibar to the magnitudes close to the initial pressure supersaturation. Nevertheless, there is a general tendency for the temperature effect on the formation of abnormal growth. Such transitions at 0.83 K are not observed either at the growth stage after nucleation in a metastable liquid, or during the subsequent growth of the crystal for longer than 1 s (Fig. 2). As it concerns temperature 0.73 K, the crystal growth occurs free from abnormal state of the crystal facets for 0.4 s if the supersaturation of crystal nucleation does not exceed $\sim 2$ mbar  (Fig. 3a). Increasing the initial supersaturation to $\sim 4$ mbar results in observing two pressure jumps at the crystal growth stage. The pressure jump at 0.66 K is observed as well as on the pressure records during the growth of the crystal nucleated at supersaturation $\sim 2$ mbar  (Fig. 4a). For the lower temperatures, we detect the numerous pressure jumps on the $p_1(t)$  records of crystal growth with the same initial supersaturation, see Figs. 5a and 6a.

To summarize, the higher temperatures prevent from the formation of an anomalous state at the helium crystal facets. The boundary which approximately divides the $\delta p_0 - T$ plane  into the normal and abnormal growth regions reflects this fact in kind. The question how sharply the temperature suppresses the formation of the burst-like growth regime in the range 0.73 - 0.83 K remains open.

\subsection{The abrupt nature of the transition from the fast kinetics of facet growth to the normal one.}
Section 2 presents a criterion for the moment of the fast-to-slow transition in the crystal facet kinetics. In the lack of visual control of the crystal shape, one can only assert that the fast crystal growth is blocked by the \textit{c} facets and at least by three \textit{a} facets at angle 120$^\circ$
at the moment in the derivative $dp_1/dt$ discontinuity.

The fastest changes in the derivative $dp_1/dt$ are observed in the quasi periodic and stochastic sequences of pressure jumps with the small amplitude and duration, see plots 4--7. The negative value of derivative $dp_1/dt$ in the burst-like growth state is replaced with the positive one. The typical duration of such changes is one or two measurement points, that is, no more than 80-160 $\mu$s.

The $p_1(t)$ dependence in a sequence of long alternating anomalous and normal states of the crystal facets differs from those considered above (Fig. 7, upper plot). The transition of the crystal facets to the burst-like growth mode results in the rapid growth of the crystal. This, in turn, induces the pressure oscillations inside the container.
This problem is discussed in Ref.\cite{VLT2001} in detail. According to the result of calculations, the occurrence of oscillations is determined by the value $K$, which should exceed the magnitude 2-5~s/m. The surface growth rate is also determined by supersaturation. As follows from the measurements\cite{UFN}, this velocity is much less than the speed of sound in a liquid and does not exceed 5m/s.
The average pressure magnitudes during this time are close to the phase equilibrium pressure. The end of oscillations is accompanied by a drastic increase in pressure,  indicating the transition of the crystal facets to the normal state with the slow kinetics. The estimate of the transition time interval yields the magnitude shorter than 200 $\mu$s.

Note that this magnitude is close to the time of the normal-to-anomalous growth transition studied earlier. The video shows that the transition time to the fast growth state for all crystal facets does not exceed 40 $ \mu$s, Fig.12 \cite{UFN}. The change in the facet kinetics from anomalous to normal and vice versa occurs for the time no longer than tens of microsecond.  This is significantly smaller as compared with the time of changing the external factors such as temperature and fluid flow into the container. From this point of view the phase transition in the crystal facet kinetics has a discontinuous character.

\subsection{Alternative explanation for the ultraslow growth of the defect-free \textit{c} facet.}
In experiments \cite{RF-1, RF-2} it is found with the aid of optical method that before the start of the burst-like growth mode, the defect-free \textit{c}-facet grows at a very low rate $(1-5)\cdot10^{-8}$ cm/s. Taking the data scatter into account, we see that the mobility of the crystal facet is constant in the range 2--20mK and then starts to decrease as the temperature increases. The theoretical explanation of this effect is proposed in paper \cite{AM} where the vacancy mechanism of crystal growth is considered.

The discovery of the crystal growth regime accompanied by the pressure jumps of small amplitude and duration makes it possible to propose an alternative explanation for the ultraslow growth of the defect-free \textit{c} facet. The short-term transition of the crystal \textit{c} facet to the burst-like growth state corresponding to such facet growth rate produces a small pressure jump. The  pressure meter with a long time constant  averages the pressure jumps out. This is not visible on the $p(t)$ record. The optical technique based on digitizing the CCD camera image is inertial as well and does not register any single tiny shift of the crystal surface. Then, the ultraslow \textit{c} facet growth observed results  from averaging the sequence of small facet growths during its short-term transition to the burst-like growth state.

Within the framework of this model the ultraslow growth is determined by the same physical mechanism of unknown nature which the burst-like growth mode demonstrates at high pressure supersaturations. From the plots of the cumulative pressure distribution of burst-like growth instability \cite{RF-2} (Fig.12) and their approximation, it follows that the probability of small amplitude pressure jumps is exponentially small. The transition of the crystal facet to abnormal state is rare. The growth in  such single transition is small.

The model is qualitatively confirmed by comparing the temperature variation for the cumulative distribution of  the pressure burst-like growth instability \cite{RF-2} (Fig. 13) with the temperature behavior of the ultraslow growth rate (Fig. 11). As the temperature increases, the median of the distribution shifts to  higher supersaturations. Then, according to the logic of the model proposed, the probability reduces for appearing the short-term low-amplitude transitions at this distribution tail. Hence, it follows that the ultraslow growth rate  should decrease as the  temperature grows. This conclusion is consistent with the experimental data demonstrating the  reduction in the ultraslow growth rate as the temperature increases.

\section{Conclusion.}
The development of the experimental technique has significantly extended the prospects for studying the mysterious phenomenon, namely,  drastic change in the crystal facet growth rate by several orders of the magnitude. All the previous studies are carried out for a short time from the moment of crystal nucleation in the metastable liquid to the completion of crystal growth, that is, for $\sim$ 0.5 --1 ms. At present, in one experiment the numerous transitions of the crystal facet growth kinetics from normal to anomalous and vice versa are observed, providing us an opportunity to extend the research methods. This result is partially facilitated by the fact that the  mobility of crystal facets in the normal state has a minimum in the range 0.5 -- 0.8 K. Therefore, the change in the crystal facet kinetics becomes readily distinguishable.

An important qualitative result is setting the fact that the reverse transition from the fast to slow facet kinetics occurs simultaneously or with an insignificant delay. So far, it has been known that the kinetics becomes normal in  $\sim$0.1 s after the growth of the crystal from the metastable liquid in the burst-like growth mode. Currently, the conditions imposed on the physical mechanism for the transition in the growth kinetics are determined. One should look for a phenomenon that can induce the fast facet growth kinetics simultaneously or with a slight delay and stop it as well.

Our measurements demonstrate  a variety of crystal growth regimes   with switching the kinetics. The quasi periodic pressure jumps are observed  in the kinetics without the drops to the phase equilibrium pressure, stochastically occurring transitions, formation of a sequence of the long time periods of the normal growth states resulting in the long-term states with the high growth kinetics. So far, it has not been possible unambiguously to link the onset of certain regimes with the magnitude of supersaturation, previous history of crystal growth, growth rate, or crystal size.

The discovery of  new quality, namely, drastic return of the crystal facet growth kinetics to the normal one, opens  new field of research in the mysterious burst-like crystal growth phenomenon.

\begin{acknowledgements}
The author is grateful to V.~V.~Dmitriev for the possibility of performing these experiments at Kapitza Institute for Physical Problems RAS. The author is also grateful to V.~V.~Zavyalov for supporting this work, S.N.Burmistrov for helpful comments and V.~S.~Kruglov for interest to the work.
\end{acknowledgements}

\vspace{6pt}
\appendixname{ I}

Let us consider the spherical crystal growth in the closed volume at pressure supersaturation $\delta p_0$. In this case the rate $\dot {R} = V_{surf}$ from equation (\ref{Eq7}) and the parameter $\alpha$ should be replaced by $2\alpha$. The first equation of system (2) and equation (\ref{Eq7}) describe the crystal growth taking the curvature of crystal  surface into account. The growth stops when an additional pressure $p_{stop}$ due to the surface curvature equals the liquid overpressure. This moment is shown by the arrow B on the experimental record in Fig. 2. In Fig. 10 the dependences are shown for a ratio $p_{stop}/p_0$ as a function of the initial overpressure obtained from the calculation and experimental data.
\begin{figure}
\begin{center}
\includegraphics[%
  width=0.65\linewidth,
  keepaspectratio]{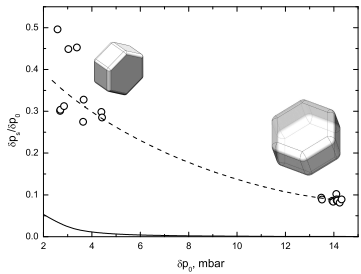}
\end{center}
\caption{The ratio of the pressure of crystal growth deceleration to the initial supersaturation at 0.83 K.}
\label{fig10}
\end{figure}
All experimental points lie above the curve calculated. This means that the edge curvature is larger than that obtained with calculating for the spherical crystal. Using pressure $p_{stop}$, we calculate the curvature of the crystal edges and reconstruct the crystal shape. It is assumed that the crystal shape is correct and the ratio of the growth rates of the basic and lateral facets equals 1:2.

\vspace{6pt}
\appendixname{ II}

Let us consider  container with a liquid and connected to the external system filled with gas at room temperature. We have also the capillary of constant cross-section and the constant mass of helium in it. When mass $\Delta m$ flows through the capillary into the container, the pressure in the container increases by $\delta p_1$ while in the external system it decreases by $\delta p_2$. The ratio of the restoring forces in the given geometry is $\frac{\delta p_1}{\delta p_2} \cong -16$. Thus, the main restoring force is generated by compressing the liquid in the container. Neglecting the response from the external system, we obtain the equation for velocity $v$ of oscillations in the capillary
\begin{equation}
\rho S L_{eff} \frac{d^2 v}{dt^2} + \rho S^2 Z \frac{dv}{dt} + \frac{S^2}{k_L V_1}v = 0.
\label{EqA2}
\end{equation}
Here $S$ is the cross-sectional area of the capillary and $L_{eff}$ is its effective length.  The parameters $Z$ and  $L_{eff}$ are determined by the frequency and attenuation of oscillations. In the containers with the large internal volume, as can be seen from the equation, the role of energy dissipation enhances and the solutions become of relaxation type. For this reason, such oscillations are not observed in the experiments with the optical containers of volume 4 - 10 cm$^3$.

\end{document}